\documentclass[doublecol]{epl2}
\usepackage{graphicx}
\usepackage{dcolumn}
\usepackage{bm}
\usepackage{amssymb}
\title{Atomistic mechanism of physical ageing in glassy materials}

\author{Mya Warren \and J{\"o}rg Rottler}

\institute{Department of Physics and Astronomy, The University of
British Columbia, 6224 Agricultural Road, Vancouver, BC, V6T 1Z1,
Canada}


\abstract{ Using molecular simulations, we identify microscopic
  relaxation events of individual particles in ageing structural
  glasses, and determine the full distribution of relaxation times.
  We find that the memory of the waiting time $t_w$ elapsed since the
  quench extends only up to the first relaxation event, while the
  distribution of all subsequent relaxation times (persistence times)
  follows a power law completely independent of history. Our results
  are in remarkable agreement with the well known phenomenological
  trap model of ageing. A continuous time random walk (CTRW)
  parametrized with the atomistic distributions captures the entire
  bulk diffusion behavior and explains the apparent scaling of the
  relaxation dynamics with $t_w$ during ageing, as well as observed
  deviations from perfect scaling.}

\begin{document}

\maketitle 

The dynamics of glass forming materials is characterized by slow,
spatially heterogeneous and temporally intermittent processes
\cite{Houches_book,Ediger_ARevPhysChem,Kob_PRL79,shintani2006,Vollmayr_PRE72,Weeks_JCM15}.
Long periods of caged particle vibrations are interrupted by rapid,
localized structural relaxations. Dynamical heterogeneity near the
glass transition leads to stretched exponential relaxation of
correlation and response functions, as well as subdiffusive,
non-Gaussian transport. In the glassy state, configurational degrees
of freedom continue to evolve slowly. This phenomenon is called
physical ageing \cite{Struik} and is a defining and universal feature
of nonequilibrium glassy dynamics \cite{cipelletti2005}.  Structural
relaxations become increasingly sluggish with the waiting time $t_w$
elapsed since vitrification, and almost all material properties change
as a result.

Ageing often manifests itself as a simple rescaling of the slow,
configurational part of the global correlation functions $C(t,t_w) =
C_{vib}(t)+C_{conf}(t/t_w^{\mu})$, where $\mu$ is the ageing
exponent. This scaling behaviour has been observed in a wide range of
disordered materials \cite{Houches_book} and is particularly
pronounced in polymer glasses \cite{Struik}, colloidal suspensions
\cite{wang2006,martinez2008}, and physical gels \cite{Cloitre_PRL85}.
Recent simulations of structural glasses show that scaling also
applies approximately to the distribution of local correlation
functions \cite{Castillo_NaturePhys3} and the distribution of particle
displacements (van Hove function) \cite{Warren_PRE07}, which
superimpose when plotted at times of equal mean. Scaling of local
correlations can be proved to hold exactly for certain spin glass
models whose dynamical action obeys time reparametrization invariance
\cite{Castillo_PRB68}. Despite the apparently universal nature of this
phenomenon \cite{Castillo_PRL09}, the physical mechanism behind it
remains elusive.

In this Letter, we directly observe the atomistic dynamics underlying
physical ageing on a single particle level through molecular dynamics
simulations of two model systems for atomic and polymeric glasses.
We consider both the well-known amorphous 80/20 binary Lennard-Jones
mixture \cite{Kob_PRE51}, as well as a bead-spring model for polymer
glasses \cite{Binder_PRE57,Baschnagel_JPCM12}, where particles
interact via the Lennard-Jones potential and are coupled together by
stiff springs to form chains of length 10. Twenty thousand particles
each are simulated in a periodic simulation box. For the polymer
(80/20) model, the initial ensemble is equilibrated at $T=1.2$ (4.0),
and then quenched rapidly to $T=0.25$ (0.33) to form a glass.  The
system is aged at constant pressure for various waiting times $t_w$
and then the dynamics are followed as a function of the measurement
time $t$. All results are quoted in Lennard-Jones units.

\begin{figure}[t]
\begin{centering}
\includegraphics[width=8cm]{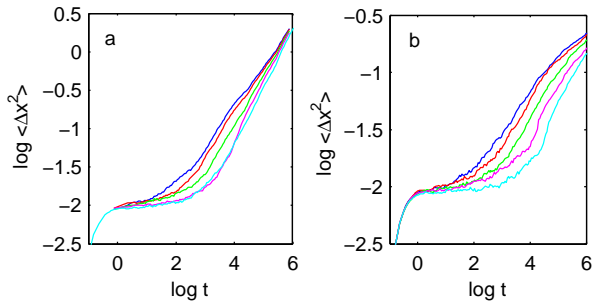}
	\caption{Mean squared displacement for (a) the binary
          Lennard-Jones mixture and (b) the polymer model for waiting
          times $t_w=750$, $2250$, $7500$, $22500$, and $75000$ (left
          to right). From superposition in the cage escape regime we
          deduce $\mu=0.51$ for the Lennard-Jones mixture and
          $\mu=0.62$ for the polymer model. In the polymer model the
          terminal slope is smaller due to Rouse dynamics.}
	\label{fig:msd}
\end{centering}
\end{figure}

In both models, the effect of ageing can be immediately observed
through the one-dimensional mean squared displacement $\langle \Delta
x^2(t,t_w) \rangle$ shown in Fig.~\ref{fig:msd}, which exhibits three
regimes: a short time vibrational regime precedes a long plateau
characterized by ``caged'' motion, followed by a cage escape or
$\alpha$-relaxation regime.  The cage escape time increases with
increasing $t_w$, but $\langle \Delta x(t,t_w)^2 \rangle$-curves for
different waiting times can be superimposed in the cage escape regime
by rescaling time with $t_w^{\mu}$. At long times $t\gg t_w$, scaling
of the mean squared displacement fails for both models, and curves for
the different ages merge.  Experimental creep compliance curves show
the same behavior, which is a well-known consequence of subageing
($\mu<1$) \cite{Struik}.

We proceed to analyze the intermittent dynamics in these systems by
decomposing single particle trajectories into discrete hopping events.
To investigate the hopping dynamics, a subset of five thousand
particle trajectories are recorded.  A running average of the atom
position and its standard deviation is found for each atom over a
small time window of 400 (40 individual snapshots of the particle
positions) \cite{comm1}. A hop is identified when the standard
deviation $\sigma$ is greater than a threshold of $2 \langle \sigma
\rangle$ \cite{comm2}.  This method differs from the usual technique
of identifying hops via a threshold in the hop distance
\cite{Vollmayr_PRE72,Chandler_JCP127,Chaudhuri_JPCM}, and effectively
captures hops at all length scales through their increased activity
during the relaxation event. While the correlation factor between hops
is nearly zero, approximately $10\%$ of hops are either forward or
backward correlated.  Removing these hops does not affect the results.
 
\begin{figure}[t]
\begin{centering}
\includegraphics[width=8cm]{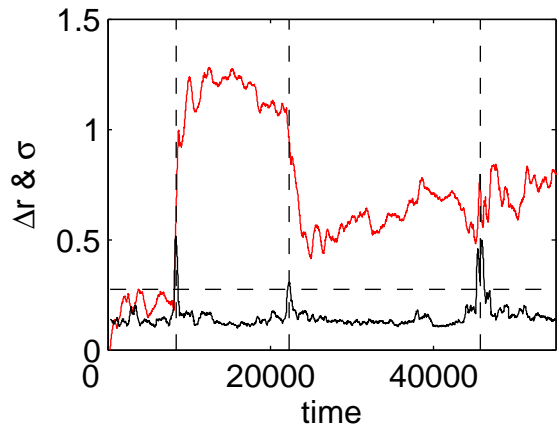}
	\caption{Displacement (red/gray) and standard deviation
          (black) from the running average of a typical atomic
          trajectory. The threshold in the standard deviation
          used for identifying hops is indicated by a horizontal
          dashed line. The times where hops are identified are
          indicated by vertical dashed lines.}
	\label{fig:trajectory}
\end{centering}
\end{figure}

\begin{figure*}[t]
\begin{centering}
\includegraphics[width=17.5cm]{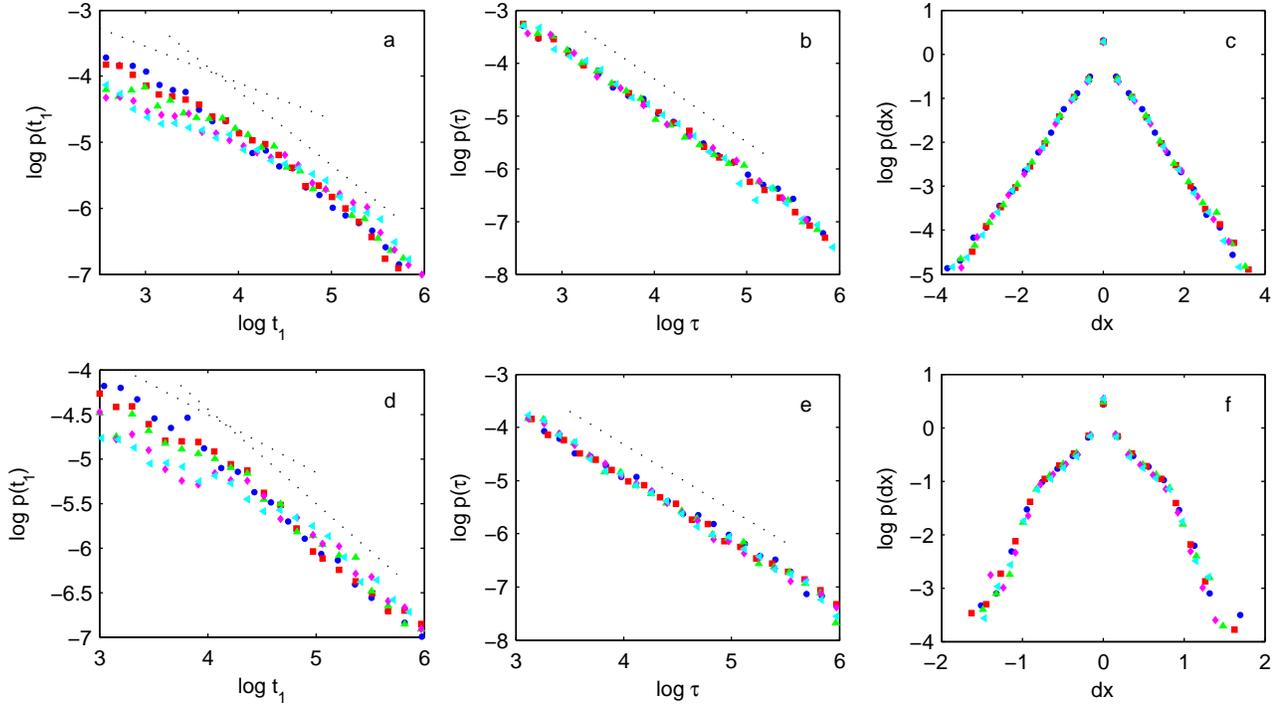}
	\caption{Distributions of first hop time $t_1$, persistence
          time $\tau$, and displacement $dx$ for the binary
          Lennard-Jones mixture (a-c) and the polymer model (d-f) for
          waiting times $t_w=750$ ($\bigcirc$), $2250$ ($\square$),
          $7500$ ($\triangle$), $22500$ ($\lozenge$), and $75000$
          ($\lhd$). Dotted lines indicate power laws and have slopes
          of -0.55 and -1.1 (a), -1.20 (b), -0.65 and -1.1 (d), and
          -1.23 (e).}
	\label{fig:p}
\end{centering}
\end{figure*}

Fig.~\ref{fig:trajectory} shows a typical atomic trajectory with the
hops identified using this criterion.  The initial hop times $t_1$,
the times between all subsequent hops $\tau$, also called the
persistence time, and the displacements $dx$ were binned to form
distribution functions $p(t_1)$, $p(\tau)$, and $p(dx)$. The initial
hop time distribution is shown in Fig.~\ref{fig:p}(a) for the
Lennard-Jones mixture and in (d) for the polymer. For all waiting
times, it appears to take the form of two power laws. The exponents
are waiting time independent, but the crossover time between the two
regimes shifts towards longer times with increasing $t_w$.

Figures \ref{fig:p}(b) and (e) show that the distributions $p(\tau)$
also decay as a power law: $p(\tau) \sim \tau^{-1.20}$ (Lennard-Jones
mixture) and $p(\tau) \sim \tau^{-1.23}$ (polymer). By contrast,
$p(\tau)$ is independent of $t_w$ and shows no ageing effects. The
one-dimensional hop displacement distribution (Fig.~\ref{fig:p}(c) and
(f)) is sharply peaked at zero displacement, and is also independent
of waiting time. This is perhaps not surprising, as it has been shown
that the density and the dynamical correlation lengths vary little
during ageing \cite{Weeks_JCM15}.  $p(dx)$ has a purely exponential
form for the mixture, whereas in the polymer model it decays more
rapidly in the tails due to chain connectivity effects. Both the first
hop time distribution, and the displacement distribution are in
qualitative agreement with results obtained by analyzing hops between
adjacent inherent structures
\cite{Heuer_PRL91,Heuer_PRE78,Heuer_JPCM20}.

The measured distributions have much in common with the well-known
phenomenological trap model of ageing \cite{Bouchaud_JPA}. In the trap
model, as in our results, $p(\tau)$ is constant and $p(t_1)$
continuously evolves towards longer times in the ageing regime.  In
fact, even the form of our measured distributions is similar. The
distributions $p(\tau)$ and $p(t_1)$ decay as weak power laws at long
times with an exponent $-(1+x)$, where $x<1$ is the noise temperature
which governs activated hopping between states \cite{Sollich_JR44}. In
this formulation, $x$ is either a constant (exponential density of
states), or more realistically
\cite{Heuer_PRL91,Heuer_PRE78,Heuer_JPCM20}, a slowly varying function
of time (Gaussian density of states) \cite{Bouchaud_JPA}. We do not
observe any curvature in $p(\tau)$, but cannot rule out its existence
at longer timescales.  While our results clearly show subageing
behavior, the ageing exponent in the trap model is strictly unity
\cite{Bouchaud_JPA}. However, this holds true only in the limit $t_w
\rightarrow \infty$. We have observed that depending on the quench
conditions, there can be a long lived transient regime where
correlation functions approximately scale with an ageing exponent
$\mu<1$.

To confirm that our analysis captures all important physics, we
recompute the ageing dynamics through stochastic realizations of a
continuous time random walk (CTRW) using the measured hop
distributions $p(t_1,t_w)$, $p(\tau)$ and $p(dx)$ as input. For the
polymer model, a pure random walk is insufficient as the chain
connectivity induces subdiffusive Rouse dynamics. The mean squared
displacement of the atoms as a function of the number of hops $n$ is
found to be $\langle \Delta x^2 \rangle \sim n^{1/2}$ ($\langle \Delta
x^2 \rangle \sim n$ for the binary mixture). In the CTRW, the Rouse
behaviour is accounted for by choosing a new particle position from a
Gaussian distribution widening as $n^{1/2}$ at every step. This
procedure creates the proper displacement statistics. In both systems,
the initial displacement is chosen such that the mean squared
displacement coincides with the MD data at the shortest measured
relaxation time. This is slightly higher than the actual vibrational
amplitude, but accounts for relaxations which occur at times down to
the vibrational timescale, but could not be captured otherwise. 

\begin{figure}[t]
\begin{centering}
\includegraphics[width=8cm]{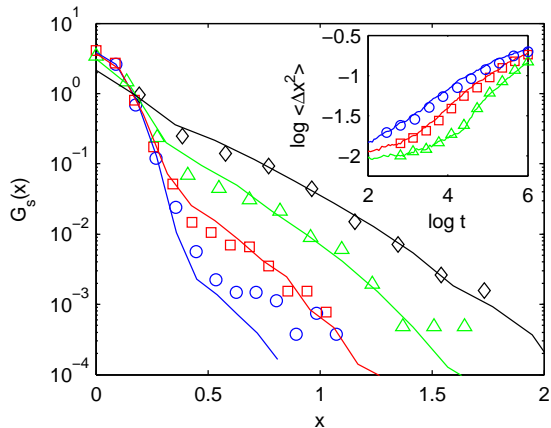}
	\caption{Van Hove function for the polymer model at $t_w =
          75000$, and times of 400, 4000, 40000, 400000 (left to
          right). Solid lines are molecular dynamics data and symbols
          are the results of the CTRW parameterized with the measured
          distributions. Inset shows the agreement between the mean
          squared displacement from MD (lines) and CTRW (symbols) at
          $t_w$= 750, 7500, and 75000 (left to right).}
	\label{fig:CTRW}
\end{centering}
\end{figure}

The full distribution of displacements (van Hove function)
\begin{equation}
	G_s(x,t,t_w)=\langle \delta \left( x - \left|x_i (t,t_w) - x_i
        (0,t_w) \right| \right) \rangle
\end{equation}
as well the as the mean squared displacements are compared with MD
data in Fig.~\ref{fig:CTRW} for the polymer model. As time progresses,
a Gaussian caged peak at small displacements loses particles to
widening exponential tails.  The origin of the exponential tails has
recently generated intense interest, as they appear to be a universal
feature of structural glasses
\cite{Kob_PRL99,Chaudhuri_JPCM,Langer_PRE77,Langer_PRE78,Schweizer_PRE77}. In
our calculation, they arise naturally from the wide distribution of
persistence times.  The CTRW reproduces the true ageing dynamics
exceedingly well, with no fit parameters. The same conclusion holds
for the binary mixture. As in the trap model, the power law form of
$p(\tau)$ is sufficient to generate ageing. Distributions $p(t_1,t_w)$
evolved via the CTRW from the shortest waiting time alone closely
track measured distributions. Ageing emerges here as a
self-generating, dynamical phenomenon.

Our results clearly show that there are two relevant timescales in the
relaxation dynamics of the glass: the first hop time which
experiences ageing, and the subsequent hop times which are $t_w$
independent. This would seem to refute the hypothesis that the
dynamics are universally reparametrized with waiting time as has been
suggested for spin glasses \cite{Castillo_NaturePhys3}, and may serve
to explain some deviations from local dynamical scaling that have been
observed. In ref.~\cite{Castillo_NaturePhys3}, the authors note that
the distribution of local correlation functions for a binary
Lennard-Jones glass superimposes to first order at times where the
global correlation function is equal. However, good collapse was found
only for large and small global correlation with significant
deviations for $C=0.3$ and $C=0.5$.

\begin{figure}[t]
\begin{centering}
\includegraphics[width=8cm]{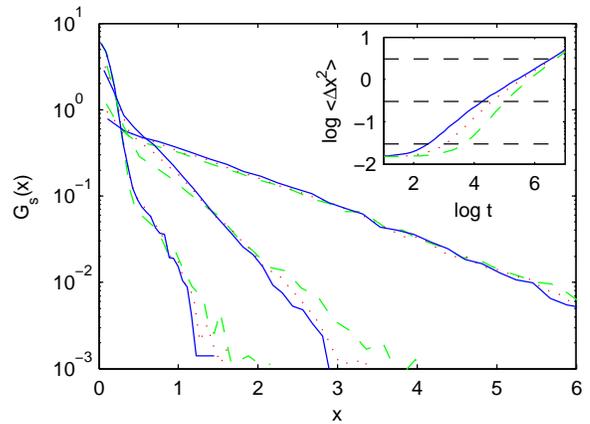}
	\caption{Van Hove functions from ageing CTRWs at three
          different waiting times $750$ (solid), $7500$ (dotted), and
          $75000$ (dashed) plotted at times $t$ where $\langle \Delta
          x^2 \rangle =$ 0.03, 0.3, and 3 as indicated by horizontal
          lines in the inset. For these values, the correlation
          function $C(t,t_w)=\langle exp(7.2 i [x(t_w+t)-x(t_w)])
          \rangle \approx$ 0.65, 0.3, and 0.1. Data is generated from
          the distributions $p(t_1,t_w=0)=t_1^{-1.2}$ and
          $p(\tau)=\tau^{-1.2}$.}
	\label{fig:CTRW-superpos}
\end{centering}
\end{figure}

We further investigate the local dynamical scaling by using the CTRW
model to calculate the van Hove distribution for three waiting times
and three different values of the mean in
Fig.~\ref{fig:CTRW-superpos}. Note that here $p(t_1,t_w)$ is generated
by evolving the system directly in the ageing CTRW. The superposition
of the van Hove functions displays the same behavior reported in
ref.~\cite{Castillo_NaturePhys3}. At short times (large $C$), the van
Hove functions closely superimpose at times of equal mean. In this
regime, few atoms have hopped and their mean squared displacement and
van Hove function are dominated by the distribution $p(t_1,t_w)$,
which scales with $t_w$. For the mid-range of the correlation
functions, the overlap of the van Hove functions becomes weaker. The
long waiting time distributions have a sharper trapped peak and wider
tails than the short $t_w$ function. This is due to the fact that the
width of the tails is controlled by $p(\tau)$, whereas the height of
the caged peak is controlled by $p(t_1,t_w)$. At very long times $t
\gg t_w$ (small $C$), the mean number of hops per atom is much greater
than one, and the waiting time independent behavior begins to dominate
the dynamics. In this case, the mean squared displacement curves merge
and there is trivial superposition of the van Hove functions.

To summarize, we have presented a complete characterization of
particle diffusion in ageing structural glasses. We find that on a
microscopic level, time is not simply reparameterized during
ageing. Only the first hop time depends on the waiting time, and the
particle subsequently ``forgets'' its age. The evolution of the first
hop time distributions is remarkably close to that assumed in the trap
model with annealed disorder. Although this model is a rather abstract
picture of glassy dynamics, our results indicate that it may be closer
to physical reality on a single particle level than previously
thought. An explicit mapping of the MD data onto a CTRW with measured
distributions of hop times and displacements is able to completely
reproduce all dynamical features of the relaxing glass, and provides
an explanation for the scaling with $t_w$ and small deviations
thereof.

We thank the Natural Sciences and Engineering Research Council of
Canada (NSERC) and the Canadian Foundation for Innovation (CFI) for
financial support.


\begin{thebibliography}{10}
\bibitem{comm1}
The choice of time window affects the width of the resulting displacement
  distribution and the normalization constant of the persistence time
  distribution, but the slope of the persistence time and the distribution of
  first hop times are unaffected.
\bibitem{comm2}
Increasing the threshold broadens both hop time and displacement distributions;
  however, thresholds in a range of 1.5 to 2.5 give self-consistent results for
  the mean-squared displacement in the continuous time random walk. A threshold
  of two was chosen as a compromise between resolving ageing and reducing
  noise.
\end{thebibliography}

\end{document}